\documentclass[onecolumn]{aa_modified}

\usepackage{allrunes}
\usepackage{graphicx}
\usepackage{txfonts}
\usepackage{wasysym}
\usepackage{array}
\usepackage[pdftitle={Exoplanet Vision 2050}, colorlinks = true, breaklinks = true, citecolor = blue, linkcolor = blue, urlcolor = blue, pdfauthor = {Ren\'{e} Heller}]{hyperref}
\usepackage{esvect}  

\begin{document}

\title{Exoplanet Vision 2050}

\titlerunning{Exoplanet Vision 2050}
\author{Ren\'{e} Heller\inst{1}
\and
L{\'a}szl{\'o} L. Kiss\inst{2}
          }

   \institute{Max Planck Institute for Solar System Research, Justus-von-Liebig-Weg 3, 37077 G\"ottingen, Germany\\
   \email{heller@mps.mpg.de}
   \and
   Konkoly Observatory, Research Centre for Astronomy and Earth Sciences, Hungarian Academy of Sciences, 1121 Budapest, Konkoly Thege M. {\'u}t 15–17, Hungary
\\
             }

\abstract{Is there any hope for us to draw a plausible picture of the future of exoplanet research? Here we extrapolate from the first 25 years of exoplanet discovery into the year 2050. If the power law for the cumulative exoplanet count continues, then almost 100,000,000 exoplanets would be known by 2050. Although this number sounds ridiculously large, we find that the power law could plausibly continue until at least as far as 2030, when Gaia and WFIRST will have discovered on the order of 100,000 exoplanets. After an early era of radial velocity detection, we are now in the transit era, which might be followed by a transit \& astrometry era dominated by the WFIRST and Gaia missions. And then? Maybe more is not better. A small and informal survey among astronomers at the ``Exoplanet Vision 2050'' workshop in Budapest suggests that astrobiological topics might influence the future of exoplanet research.}


\maketitle

\section{About the science and fiction of technological predictions}

Can we make a plausible and credible prediction of what humans will know about extrasolar planets in the year 2050? Such a prediction must necessarily invoke a decent amount of fiction. As scientists, we might thus end up in the realm of science fiction. But maybe we can obtain a better feeling for the plausibility of our predictions by looking at the history of astronomy-related science fiction literature.

One popular science fiction book that connected to modern astronomy and space science is the story ``De la terre {\'a} la lune'' by Jules \citet{Verne}, which pictures the first manned space mission to the Moon. Such a mission to the Moon was indeed achieved a little more than 100 years later by the Apollo 8 mission. Though many aspects of Verne's novel were impossible from a physical point of view such as the fact that the astronauts survived the launch in a capsule that was essentially a projectile accelerated by a gun powder explosion in an enormously large cannon, the general picture was realistic to some extent.

Another prominent example is Fritz Lang's 1927 movie ``Metropolis'', which plays in a dystopic future. In one scene of the movie, we see a fleet of biplanes flying through a landscape of skyscrapers. And even though biplanes have not gained acceptance as the prevailing means of transportation today, we have witnessed a strong individualization of the means of transport and in particular of commercial flight.

More on the astronomical side of science fiction, \citet{1952Obs....72..199S} argued for a dedicated spectroscopic survey of nearby sun-like stars for radial velocity (RV) shits caused by Jupiter-mass planets in very close orbits around their stars. The existence of this type of planet, nowadays referred to as ``hot Jupiter'', was only proven observationally more than 40 years later by the discovery of 51\,Peg\,b \citep{1995Natur.378..355M}. Struve's note that ``There would, of course, also be eclipses'' was also proven observationally almost 50 years later by the transit detection of another hot Jupiter, HD209458\,b \citep{2000ApJ...529L..45C}.

In a similar way, science fiction author G.~H.~Stine, in his visionary 1973 essay ``A Program for Star Flight'', laid out a plan for humans to go interstellar on a time scale of centuries \citep{Stine1973}. He proposed that astronomers would first have to survey nearby stars for planets and that the most promising techniques for exoplanet detection would be astrometry and RVs. He also predicted that the first planet around another star would be discovered in 20 years from the time of writing -- a stunning prediction of the 1995 discovery by M.~Mayor \& D.~Queloz.

Since then, with thousands of exoplanets known and with an average of about one planet being found every day now, astronomers have entered a phase of exoplanet population studies. The number of exoplanets has become sufficiently large to infer reliable exoplanet occurrence rates \citep{2012ApJS..201...15H}, which revealed different subpopulations of planets \citep[e.g. see the identification of a radius gap near 1.5 Earth radii among short period planets;][]{2017AJ....154..109F}, raising new questions about planet formation and evolution. The question is: what next? Will more planet discoveries mean more scientific progress? The European Space Agency (ESA) has recently issued a call for white papers to direct their Voyage 2050 research program.\footnote{\href{https://www.cosmos.esa.int/web/voyage-2050}{https://www.cosmos.esa.int/web/voyage-2050}} In this context, exoplanet researchers are now consolidating their thoughts on the steps after the next, that is, after the planned launch of the European missions ``PLAnetary Transits and Oscillations of stars'' \citep[PLATO;][]{2014ExA....38..249R} in 2026 and ``Atmospheric Remote-sensing Infrared Exoplanet Large-survey'' \citep[ARIEL;][]{2018ExA....46..135T} in 2028.

Here we discuss the future of exoplanet discoveries up to the year 2050. We are aware of the uncertainties that our predictions involve and we are aware of our unawareness of possibly many other unknowns. In this sense, this paper must be taken with a large grain of salt but we hope it might nevertheless be perceived as entertaining and stimulating.

\section{Exoplanet eras and exoplanet counts}

The first 100 planets or so were all found using the RV method, which is why we can safely refer to the mid- and late 1990s as the RV era. Serendipitous ground-based transit detections motivated the installation of dedicated large-scale transit surveys, which started to deliver discoveries in the middle of the first decade of the 2000s, such as the Optical Gravitational Lensing Experiment \citep[OGLE;][]{ 2003Natur.421..507K}, the Transatlantic Exoplanet Survey \citep[TrES;][]{2004ApJ...613L.153A}, the Hungarian Automated Telescope Network \citep[HATNet;][]{2004PASP..116..266B}, and the Super Wide Angle Search for Planets \citep[SuperWASP;][]{2006PASP..118.1407P}. These surveys reached a combined planet yield rate of about a dozen per year in 2010, when the total planet count had risen to the hundreds. The success of the ``Convection, Rotation and planetary Transits'' mission \citep[CoRot;][]{2009A&A...506..411A} and of the Kepler mission \citep{2010Sci...327..977B} then further boosted the detection rates. In fact, most of the exoplanets known today have been discovered with the Kepler telescope. As of 28 Nov. 2019, the NASA Exoplanet Archive\footnote{\href{https://exoplanetarchive.ipac.caltech.edu/}{https://exoplanetarchive.ipac.caltech.edu}} counts a total of 4099 confirmed exoplanets. From an exoplanet count perspective, it is thus natural to state that today's exoplanet researchers are working in a transit era.

\begin{figure*}
\centering
\includegraphics[angle= 0,width=0.57\linewidth]{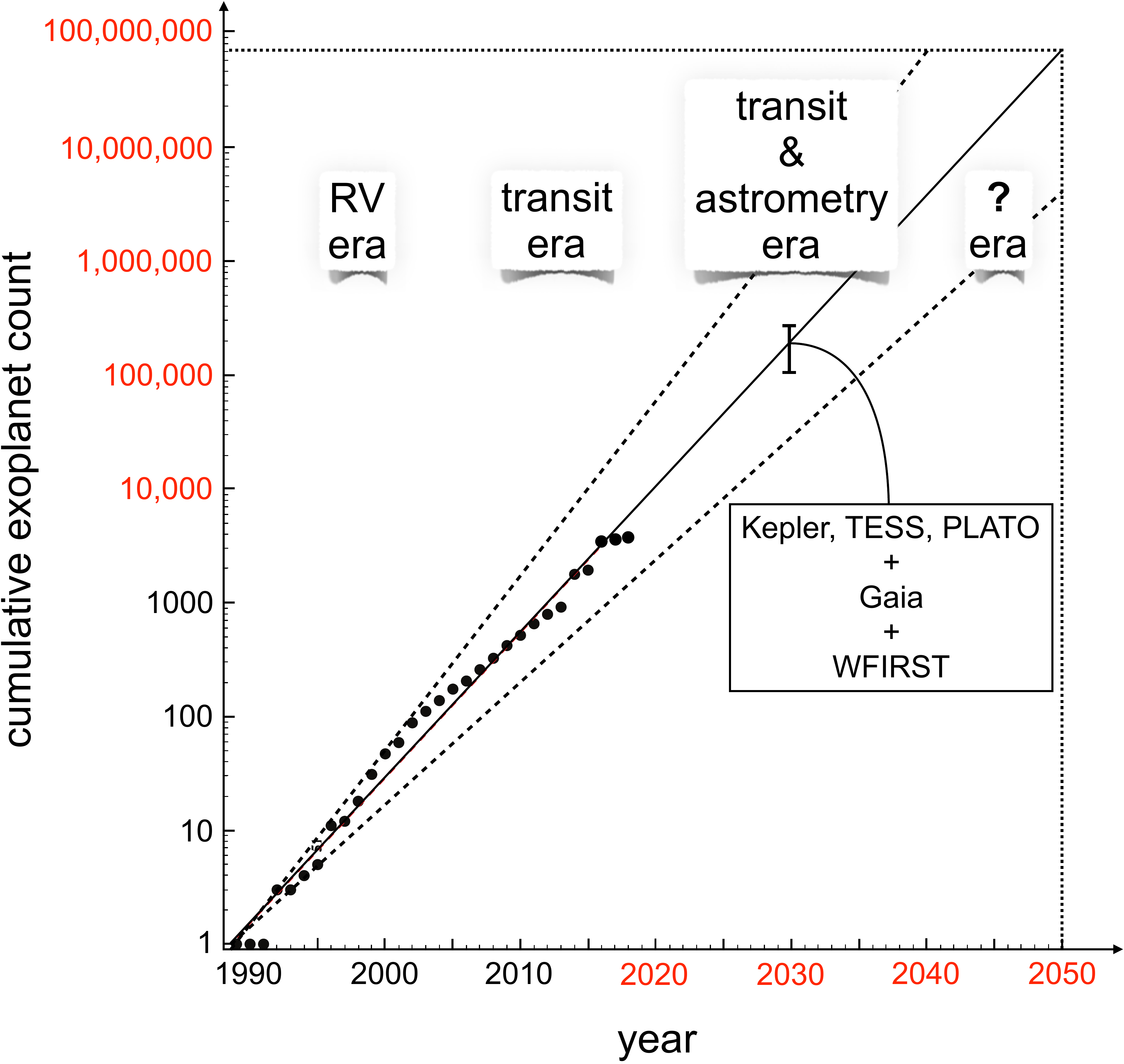}
\caption{Extrapolation of the exoplanet count into the year 2050. The black solid line is a fit to the data from 1988 to 2018, which hits a value of almost 100,000,000 in 2050. The dashed lines are the envelopes of the data. Red numbers along the abscissa and ordinate denote the ranges of extrapolation, for which no data is yet available. The vertical bar above the year 2030 refers to a combination of the currently known exoplanet population and expected exoplanet yields for TESS, PLATO, Gaia, and WFIRST as presented in the literature. The data points were taken from the \citet{NAP25187} based on data by A. Weinberger/NASA Exoplanet Archive. Adapted and reproduced with permission from the National Academy of Sciences, Courtesy of the National Academies Press, Washington, D.C.}
\label{fig:counts}
\end{figure*}

Figure~\ref{fig:counts} shows a summary of the annual exoplanet detection count as a function of time. The lower left corner of this plot contains data from the NASA Exoplanet Archive as shown by the \citet{NAP25187}.\footnote{After our presentation of Fig.\ref{fig:counts} at the ``Exoplanet Vision 2050'' workshop we have learned after that a similar plot was published online on 24 Oct. 2016 by E.~Mamajek at \href{https://twitter.com/EricMamajek/status/790786565496680449}{https://twitter.com/EricMamajek/status/790786565496680449}. He derived a power law dependence with a doubling time of 27 months.} The black solid line indicates our fit to the data, and the dotted lines show an extrapolation of the envelope of the data. The red numbers along the abscissa and the ordinate indicate the predictive part of the diagram, where we gradually leave the realm of science and enter the realm of fiction. In particular, in 2050 the fit suggests an exoplanet count of almost $10^8$, potentially corresponding to a few percent of all the planets in the Milky Way if the average planet occurrence rate would be of the order of a few planets per star. The evaluation as to how realistic such a prediction is, is left to the opinion of the reader.

That said, we can construct a plausible estimate for the year 2030. Let us start with a total lump sum of 20,000 exoplanets from the past primary Kepler mission and the completed K2 mission \citep{2014PASP..126..398H}, from today's Transiting Exoplanet Survey Satellite \citep[TESS;][]{2014SPIE.9143E..20R}, and from the planned 2026 PLATO mission \citep{2014ExA....38..249R}. Then we add the expected planet yield from the present-day Gaia mission, which ranges between 15,000 and 90,000 depending on the final mission duration, the actual occurrence rates of massive planets in wide orbits etc. \citep{2014ApJ...797...14P}. The Wide Field Infrared Survey Telescope \citep[WFIRST;][]{2015arXiv150303757S}, with an expected launch in 2025 and a mission duration of five years, is supposed to find another about 70,000 to 150,000 transiting exoplanets, details depending mostly on the actual occurrence rate of hot Jupiters in the WFIRST observing fields \citep{2017PASP..129d4401M}. In summary, our literature research suggests that a total exoplanet count of between 105,000 and 260,000 can be expected for 2030, which is indicated with a vertical error bar in Fig.~\ref{fig:counts}. Most astoundingly, we find that the power law fit goes precisely through this error bar and that this tolerance window is located entirely within the envelope of the predictions. With Gaia and WFIRST dominating the exoplanet count in the 2030s, one could reasonably predict that this will be a transit \& astrometry era of exoplanet detections.


What would be the next era, the one dominating the exoplanet population by 2050? We don't know. That said, we do have an idea when we might know. We consider two examples of an exoplanet space mission, Kepler and PLATO. The first mentioning of the photometric method as a means to find extrasolar planets goes back to \citet{1984Icar...58..121B} and the first formal proposal was rejected by NASA in 1992.\footnote{\href{https://www.elsevier.com/connect/everyone-agrees-your-method-is-not-going-to-work-william-borucki-on-the-power-of-resolve-in-nasas-kepler-mission}{elsevier.com/connect/everyone-agrees-your-method-is-not-going-to-work-william-borucki-on-the-power-of-resolve-in-nasas-kepler-mission}} Kepler launched in 2009. This implies a time from the first idea to launch of 25 years and a time from the first proposal to launch of 17 years. The case for PLATO was similar. Initially drafted in 2005 (H.~Rauer, priv. comm.), it was first proposed as a medium class mission for Europe's Cosmic Vision 2015-2025 program for launch in 2017-2018.\footnote{See the PLATO Definition Study Report (Red Book) at \href{https://sci.esa.int/s/8rPyPew}{https://sci.esa.int/s/8rPyPew}.} PLATO was rejected at this stage but ESA accepted its successor version for an M3 mission profile in 2014. The launch is now planned for 2026. This gives a time from the first proposal to (planned) launch of 19 years. If these time lines are typical for the exoplanet missions to come, then any mission that would dominate exoplanet science in 2050 would have gathered the data in the 2040s, which means that it would have been proposed in the 2020s. Thus, within the next about ten years, we might have an idea about the exoplanet searches that \textit{will have been} conducted in 2050.

\section{Future topics}

\begin{figure*}
\centering
\includegraphics[angle= 0,width=0.75\linewidth]{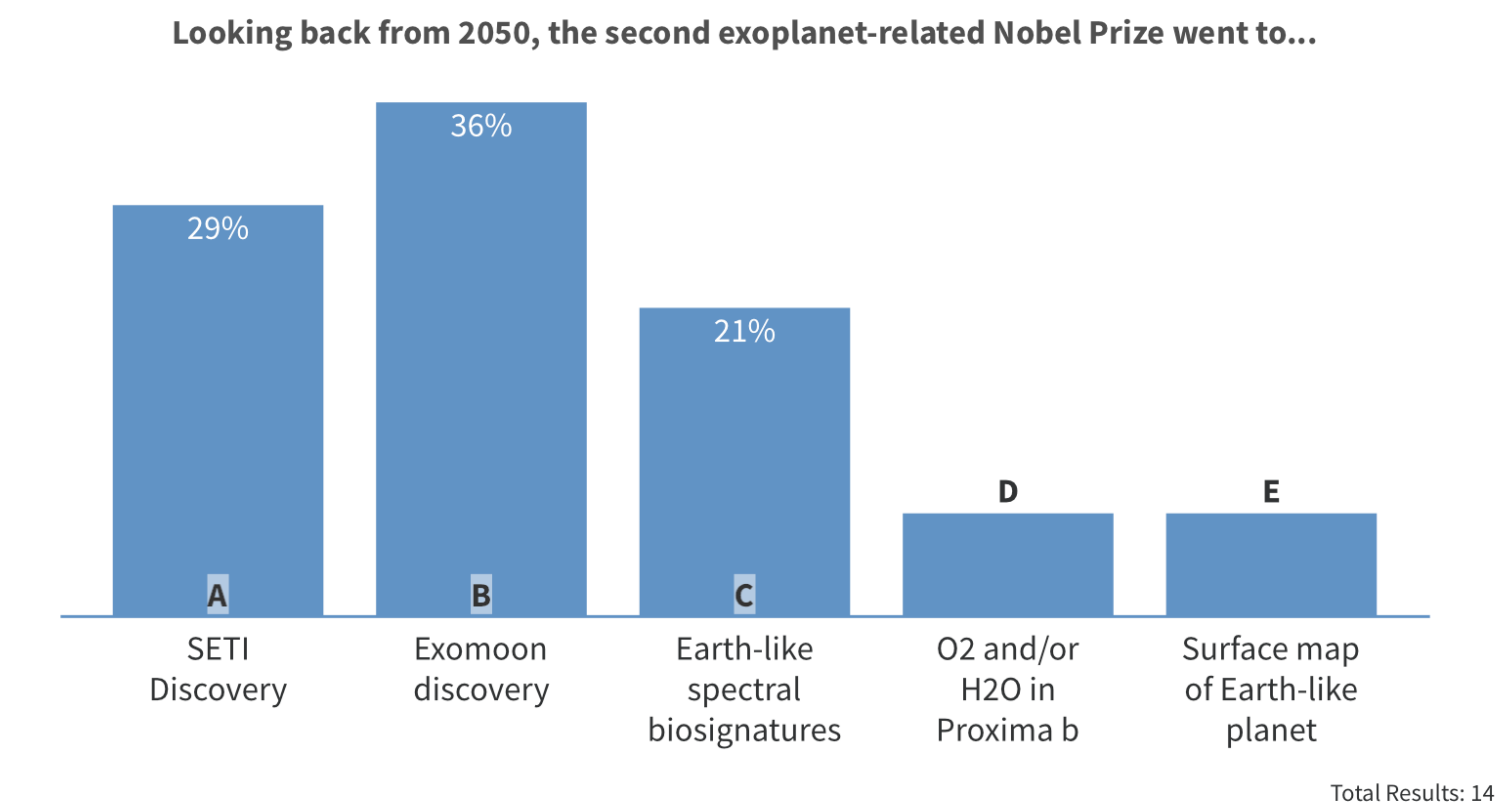}
\caption{Response of the audience at the Exoplanet Vision 2050 workshop to the question of the next exoplanet-related research topic that will be awarded with a Noble Prize}
\label{fig:NobelPrize}
\end{figure*}

More is not necessarily better. Though moving on from knowing one planet to knowing 1000 planets certainly provided the basis for almost everything that is currently known about exoplanets, it is not clear that moving on another three orders of magnitude from 1000 planets to 1,000,000 would have the same profound impact on the field. Statistics would, by definition, be better. And we would naturally know more of the individual special cases of extrasolar planets. But would it shift our paradigms about planet formation as much as the discovery of the first few hot Jupiters? Maybe it is more reasonable not to simply hunt for ever more planets but to focus the resources on a limited amount of topics in the field of extrasolar planets that are interesting and that affect our knowledge and inspiration the most. What could these topics be?

In the year 2019, the year of the first Nobel Prize ever awarded to a discovery in the field of extrasolar planets, it is thus only natural to ask ourselves: what could be the next exoplanet-related achievement or discovery that will be awarded with a Nobel Prize? During the ``Exoplanet Vision 2050'' workshop at Konkoly Observatory in Budapest (20--22 Nov. 2019), the lead author of this report gave a talk, in which he held a vote among the audience on five proposed topics. These topics had been collected from the participants at the conference prior to the talk. Figure~\ref{fig:NobelPrize} shows the outcome of this live poll. The winner is an exomoon discovery, a topic that has certainly gained considerable attraction in the exoplanet community over the past decade \citep[for a review see][]{2018haex.bookE..35H}. Of equal interest to us is the range of topics that was proposed, reaching from the surface mapping of an Earth-like planet and the detection of atmospheric spectral biomarkers in an Earth-like planet or on Proxima\,b \citep{2016Natur.536..437A} to the discovery of extraterrestrial intelligence. In summary, our audience had the strong tendency to think of exoplanets in the context of astrobiology when it comes to future ground-breaking discoveries.

\begin{acknowledgements}
This research has made use of the NASA Exoplanet Archive, which is operated by the California Institute of Technology, under contract with the National Aeronautics and Space Administration under the Exoplanet Exploration Program. This work made use of NASA's ADS Bibliographic Services. RH is supported by the German space agency (Deutsches Zentrum f\"ur Luft- und Raumfahrt) under PLATO Data Center grant 50OO1501. LK gratefully acknowledges financial support from the Konkoly Observatory for the organization of the ``Exoplanet Vision 2050'' workshop.
\end{acknowledgements}

\bibliography{references}

\begin{thebibliography}{25}
\expandafter\ifx\csname natexlab\endcsname\relax\def\natexlab#1{#1}\fi

\bibitem[{{Alonso} {et~al.}(2004){Alonso}, {Brown}, {Torres}, {Latham},
  {Sozzetti}, {Mandushev}, {Belmonte}, {Charbonneau}, {Deeg}, {Dunham},
  {O'Donovan}, \& {Stefanik}}]{2004ApJ...613L.153A}
{Alonso}, R., {Brown}, T.~M., {Torres}, G., {et~al.} 2004, \apjl, 613, L153

\bibitem[{{Anglada-Escud{\'e}} {et~al.}(2016){Anglada-Escud{\'e}}, {Amado},
  {Barnes}, {Berdi{\~n}as}, {Butler}, {Coleman}, {de La Cueva}, {Dreizler},
  {Endl}, {Giesers}, {Jeffers}, {Jenkins}, {Jones}, {Kiraga}, {K{\"u}rster},
  {L{\'o}pez-Gonz{\'a}lez}, {Marvin}, {Morales}, {Morin}, {Nelson}, {Ortiz},
  {Ofir}, {Paardekooper}, {Reiners}, {Rodr{\'\i}guez},
  {Rodr{\'\i}guez-L{\'o}pez}, {Sarmiento}, {Strachan}, {Tsapras}, {Tuomi}, \&
  {Zechmeister}}]{2016Natur.536..437A}
{Anglada-Escud{\'e}}, G., {Amado}, P.~J., {Barnes}, J., {et~al.} 2016, \nat,
  536, 437

\bibitem[{{Auvergne} {et~al.}(2009){Auvergne}, {Bodin}, {Boisnard}, {Buey},
  {Chaintreuil}, {Epstein}, {Jouret}, {Lam-Trong}, {Levacher}, {Magnan},
  {Perez}, {Plasson}, {Plesseria}, {Peter}, {Steller}, {Tiph{\`e}ne}, {Baglin},
  {Agogu{\'e}}, {Appourchaux}, {Barbet}, {Beaufort}, {Bellenger}, {Berlin},
  {Bernardi}, {Blouin}, {Boumier}, {Bonneau}, {Briet}, {Butler}, {Cautain},
  {Chiavassa}, {Costes}, {Cuvilho}, {Cunha-Parro}, {de Oliveira Fialho},
  {Decaudin}, {Defise}, {Djalal}, {Docclo}, {Drummond}, {Dupuis}, {Exil},
  {Faur{\'e}}, {Gaboriaud}, {Gamet}, {Gavalda}, {Grolleau}, {Gueguen},
  {Guivarc'h}, {Guterman}, {Hasiba}, {Huntzinger}, {Hustaix}, {Imbert},
  {Jeanville}, {Johlander}, {Jorda}, {Journoud}, {Karioty}, {Kerjean},
  {Lafond}, {Lapeyrere}, {Landiech}, {Larqu{\'e}}, {Laudet}, {Le Merrer},
  {Leporati}, {Leruyet}, {Levieuge}, {Llebaria}, {Martin}, {Mazy}, {Mesnager},
  {Michel}, {Moalic}, {Monjoin}, {Naudet}, {Neukirchner}, {Nguyen-Kim},
  {Ollivier}, {Orcesi}, {Ottacher}, {Oulali}, {Parisot}, {Perruchot},
  {Piacentino}, {Pinheiro da Silva}, {Platzer}, {Pontet}, {Pradines},
  {Quentin}, {Rohbeck}, {Rolland}, {Rollenhagen}, {Romagnan}, {Russ}, {Samadi},
  {Schmidt}, {Schwartz}, {Sebbag}, {Smit}, {Sunter}, {Tello}, {Toulouse},
  {Ulmer}, {Vandermarcq}, {Vergnault}, {Wallner}, {Waultier}, \&
  {Zanatta}}]{2009A&A...506..411A}
{Auvergne}, M., {Bodin}, P., {Boisnard}, L., {et~al.} 2009, \aap, 506, 411

\bibitem[{{Bakos} {et~al.}(2004){Bakos}, {Noyes}, {Kov{\'a}cs}, {Stanek},
  {Sasselov}, \& {Domsa}}]{2004PASP..116..266B}
{Bakos}, G., {Noyes}, R.~W., {Kov{\'a}cs}, G., {et~al.} 2004, \pasp, 116, 266

\bibitem[{{Borucki} {et~al.}(2010){Borucki}, {Koch}, {Basri}, {Batalha},
  {Brown}, {Caldwell}, {Caldwell}, {Christensen-Dalsgaard}, {Cochran},
  {DeVore}, {Dunham}, {Dupree}, {Gautier}, {Geary}, {Gilliland}, {Gould},
  {Howell}, {Jenkins}, {Kondo}, {Latham}, {Marcy}, {Meibom}, {Kjeldsen},
  {Lissauer}, {Monet}, {Morrison}, {Sasselov}, {Tarter}, {Boss}, {Brownlee},
  {Owen}, {Buzasi}, {Charbonneau}, {Doyle}, {Fortney}, {Ford}, {Holman},
  {Seager}, {Steffen}, {Welsh}, {Rowe}, {Anderson}, {Buchhave}, {Ciardi},
  {Walkowicz}, {Sherry}, {Horch}, {Isaacson}, {Everett}, {Fischer}, {Torres},
  {Johnson}, {Endl}, {MacQueen}, {Bryson}, {Dotson}, {Haas}, {Kolodziejczak},
  {Van Cleve}, {Chandrasekaran}, {Twicken}, {Quintana}, {Clarke}, {Allen},
  {Li}, {Wu}, {Tenenbaum}, {Verner}, {Bruhweiler}, {Barnes}, \&
  {Prsa}}]{2010Sci...327..977B}
{Borucki}, W.~J., {Koch}, D., {Basri}, G., {et~al.} 2010, Science, 327, 977

\bibitem[{{Borucki} \& {Summers}(1984)}]{1984Icar...58..121B}
{Borucki}, W.~J. \& {Summers}, A.~L. 1984, \icarus, 58, 121

\bibitem[{{Charbonneau} {et~al.}(2000){Charbonneau}, {Brown}, {Latham}, \&
  {Mayor}}]{2000ApJ...529L..45C}
{Charbonneau}, D., {Brown}, T.~M., {Latham}, D.~W., \& {Mayor}, M. 2000, \apjl,
  529, L45

\bibitem[{{Fulton} {et~al.}(2017){Fulton}, {Petigura}, {Howard}, {Isaacson},
  {Marcy}, {Cargile}, {Hebb}, {Weiss}, {Johnson}, {Morton}, {Sinukoff},
  {Crossfield}, \& {Hirsch}}]{2017AJ....154..109F}
{Fulton}, B.~J., {Petigura}, E.~A., {Howard}, A.~W., {et~al.} 2017, \aj, 154,
  109

\bibitem[{{Heller}(2018)}]{2018haex.bookE..35H}
{Heller}, R. 2018, {Detecting and Characterizing Exomoons and Exorings}, 35

\bibitem[{{Howard} {et~al.}(2012){Howard}, {Marcy}, {Bryson}, {Jenkins},
  {Rowe}, {Batalha}, {Borucki}, {Koch}, {Dunham}, {Gautier}, {Van Cleve},
  {Cochran}, {Latham}, {Lissauer}, {Torres}, {Brown}, {Gilliland}, {Buchhave},
  {Caldwell}, {Christensen-Dalsgaard}, {Ciardi}, {Fressin}, {Haas}, {Howell},
  {Kjeldsen}, {Seager}, {Rogers}, {Sasselov}, {Steffen}, {Basri},
  {Charbonneau}, {Christiansen}, {Clarke}, {Dupree}, {Fabrycky}, {Fischer},
  {Ford}, {Fortney}, {Tarter}, {Girouard}, {Holman}, {Johnson}, {Klaus},
  {Machalek}, {Moorhead}, {Morehead}, {Ragozzine}, {Tenenbaum}, {Twicken},
  {Quinn}, {Isaacson}, {Shporer}, {Lucas}, {Walkowicz}, {Welsh}, {Boss},
  {Devore}, {Gould}, {Smith}, {Morris}, {Prsa}, {Morton}, {Still}, {Thompson},
  {Mullally}, {Endl}, \& {MacQueen}}]{2012ApJS..201...15H}
{Howard}, A.~W., {Marcy}, G.~W., {Bryson}, S.~T., {et~al.} 2012, \apjs, 201, 15

\bibitem[{{Howell} {et~al.}(2014){Howell}, {Sobeck}, {Haas}, {Still},
  {Barclay}, {Mullally}, {Troeltzsch}, {Aigrain}, {Bryson}, {Caldwell},
  {Chaplin}, {Cochran}, {Huber}, {Marcy}, {Miglio}, {Najita}, {Smith},
  {Twicken}, \& {Fortney}}]{2014PASP..126..398H}
{Howell}, S.~B., {Sobeck}, C., {Haas}, M., {et~al.} 2014, \pasp, 126, 398

\bibitem[{{Konacki} {et~al.}(2003){Konacki}, {Torres}, {Jha}, \&
  {Sasselov}}]{2003Natur.421..507K}
{Konacki}, M., {Torres}, G., {Jha}, S., \& {Sasselov}, D.~D. 2003, \nat, 421,
  507

\bibitem[{{Mayor} \& {Queloz}(1995)}]{1995Natur.378..355M}
{Mayor}, M. \& {Queloz}, D. 1995, \nat, 378, 355

\bibitem[{{Montet} {et~al.}(2017){Montet}, {Yee}, \&
  {Penny}}]{2017PASP..129d4401M}
{Montet}, B.~T., {Yee}, J.~C., \& {Penny}, M.~T. 2017, \pasp, 129, 044401

\bibitem[{National Academies~of Sciences \& Medicine(2018)}]{NAP25187}
National Academies~of Sciences, E. \& Medicine. 2018, Exoplanet Science
  Strategy (Washington, DC: The National Academies Press)

\bibitem[{{Penny} {et~al.}(2019){Penny}, {Gaudi}, {Kerins}, {Rattenbury},
  {Mao}, {Robin}, \& {Calchi Novati}}]{2019ApJS..241....3P}
{Penny}, M.~T., {Gaudi}, B.~S., {Kerins}, E., {et~al.} 2019, \apjs, 241, 3

\bibitem[{{Perryman} {et~al.}(2014){Perryman}, {Hartman}, {Bakos}, \&
  {Lindegren}}]{2014ApJ...797...14P}
{Perryman}, M., {Hartman}, J., {Bakos}, G.~{\'A}., \& {Lindegren}, L. 2014,
  \apj, 797, 14

\bibitem[{{Pollacco} {et~al.}(2006){Pollacco}, {Skillen}, {Collier Cameron},
  {Christian}, {Hellier}, {Irwin}, {Lister}, {Street}, {West}, {Anderson},
  {Clarkson}, {Deeg}, {Enoch}, {Evans}, {Fitzsimmons}, {Haswell}, {Hodgkin},
  {Horne}, {Kane}, {Keenan}, {Maxted}, {Norton}, {Osborne}, {Parley}, {Ryans},
  {Smalley}, {Wheatley}, \& {Wilson}}]{2006PASP..118.1407P}
{Pollacco}, D.~L., {Skillen}, I., {Collier Cameron}, A., {et~al.} 2006, \pasp,
  118, 1407

\bibitem[{{Rauer} {et~al.}(2014){Rauer}, {Catala}, {Aerts}, {Appourchaux},
  {Benz}, {Brandeker}, {Christensen-Dalsgaard}, {Deleuil}, {Gizon}, {Goupil},
  {G{\"u}del}, {Janot-Pacheco}, {Mas-Hesse}, {Pagano}, {Piotto}, {Pollacco},
  {Santos}, {Smith}, {Su{\'a}rez}, {Szab{\'o}}, {Udry}, {Adibekyan}, {Alibert},
  {Almenara}, {Amaro-Seoane}, {Eiff}, {Asplund}, {Antonello}, {Barnes},
  {Baudin}, {Belkacem}, {Bergemann}, {Bihain}, {Birch}, {Bonfils}, {Boisse},
  {Bonomo}, {Borsa}, {Brand {\~a}o}, {Brocato}, {Brun}, {Burleigh}, {Burston},
  {Cabrera}, {Cassisi}, {Chaplin}, {Charpinet}, {Chiappini}, {Church},
  {Csizmadia}, {Cunha}, {Damasso}, {Davies}, {Deeg}, {D{\'\i}az}, {Dreizler},
  {Dreyer}, {Eggenberger}, {Ehrenreich}, {Eigm{\"u}ller}, {Erikson}, {Farmer},
  {Feltzing}, {de Oliveira Fialho}, {Figueira}, {Forveille}, {Fridlund},
  {Garc{\'\i}a}, {Giommi}, {Giuffrida}, {Godolt}, {Gomes da Silva}, {Granzer},
  {Grenfell}, {Grotsch-Noels}, {G{\"u}nther}, {Haswell}, {Hatzes},
  {H{\'e}brard}, {Hekker}, {Helled}, {Heng}, {Jenkins}, {Johansen},
  {Khodachenko}, {Kislyakova}, {Kley}, {Kolb}, {Krivova}, {Kupka}, {Lammer},
  {Lanza}, {Lebreton}, {Magrin}, {Marcos-Arenal}, {Marrese}, {Marques},
  {Martins}, {Mathis}, {Mathur}, {Messina}, {Miglio}, {Montalban}, {Montalto},
  {Monteiro}, {Moradi}, {Moravveji}, {Mordasini}, {Morel}, {Mortier},
  {Nascimbeni}, {Nelson}, {Nielsen}, {Noack}, {Norton}, {Ofir}, {Oshagh},
  {Ouazzani}, {P{\'a}pics}, {Parro}, {Petit}, {Plez}, {Poretti}, {Quirrenbach},
  {Ragazzoni}, {Raimondo}, {Rainer}, {Reese}, {Redmer}, {Reffert},
  {Rojas-Ayala}, {Roxburgh}, {Salmon}, {Santerne}, {Schneider}, {Schou},
  {Schuh}, {Schunker}, {Silva-Valio}, {Silvotti}, {Skillen}, {Snellen}, {Sohl},
  {Sousa}, {Sozzetti}, {Stello}, {Strassmeier}, {{\v{S}}vanda}, {Szab{\'o}},
  {Tkachenko}, {Valencia}, {Van Grootel}, {Vauclair}, {Ventura}, {Wagner},
  {Walton}, {Weingrill}, {Werner}, {Wheatley}, \&
  {Zwintz}}]{2014ExA....38..249R}
{Rauer}, H., {Catala}, C., {Aerts}, C., {et~al.} 2014, Experimental Astronomy,
  38, 249

\bibitem[{{Ricker} {et~al.}(2014){Ricker}, {Winn}, {Vanderspek}, {Latham},
  {Bakos}, {Bean}, {Berta-Thompson}, {Brown}, {Buchhave}, {Butler}, {Butler},
  {Chaplin}, {Charbonneau}, {Christensen-Dalsgaard}, {Clampin}, {Deming},
  {Doty}, {De Lee}, {Dressing}, {Dunham}, {Endl}, {Fressin}, {Ge}, {Henning},
  {Holman}, {Howard}, {Ida}, {Jenkins}, {Jernigan}, {Johnson}, {Kaltenegger},
  {Kawai}, {Kjeldsen}, {Laughlin}, {Levine}, {Lin}, {Lissauer}, {MacQueen},
  {Marcy}, {McCullough}, {Morton}, {Narita}, {Paegert}, {Palle}, {Pepe},
  {Pepper}, {Quirrenbach}, {Rinehart}, {Sasselov}, {Sato}, {Seager},
  {Sozzetti}, {Stassun}, {Sullivan}, {Szentgyorgyi}, {Torres}, {Udry}, \&
  {Villasenor}}]{2014SPIE.9143E..20R}
{Ricker}, G.~R., {Winn}, J.~N., {Vanderspek}, R., {et~al.} 2014, Society of
  Photo-Optical Instrumentation Engineers (SPIE) Conference Series, Vol. 9143,
  {Transiting Exoplanet Survey Satellite (TESS)}, 914320

\bibitem[{{Spergel} {et~al.}(2015){Spergel}, {Gehrels}, {Baltay}, {Bennett},
  {Breckinridge}, {Donahue}, {Dressler}, {Gaudi}, {Greene}, {Guyon}, {Hirata},
  {Kalirai}, {Kasdin}, {Macintosh}, {Moos}, {Perlmutter}, {Postman},
  {Rauscher}, {Rhodes}, {Wang}, {Weinberg}, {Benford}, {Hudson}, {Jeong},
  {Mellier}, {Traub}, {Yamada}, {Capak}, {Colbert}, {Masters}, {Penny},
  {Savransky}, {Stern}, {Zimmerman}, {Barry}, {Bartusek}, {Carpenter}, {Cheng},
  {Content}, {Dekens}, {Demers}, {Grady}, {Jackson}, {Kuan}, {Kruk}, {Melton},
  {Nemati}, {Parvin}, {Poberezhskiy}, {Peddie}, {Ruffa}, {Wallace}, {Whipple},
  {Wollack}, \& {Zhao}}]{2015arXiv150303757S}
{Spergel}, D., {Gehrels}, N., {Baltay}, C., {et~al.} 2015, arXiv e-prints,
  arXiv:1503.03757

\bibitem[{{Stine}(1973)}]{Stine1973}
{Stine}, G.~H. 1973, Analog, 10

\bibitem[{{Struve}(1952)}]{1952Obs....72..199S}
{Struve}, O. 1952, The Observatory, 72, 199

\bibitem[{{Tinetti} {et~al.}(2018){Tinetti}, {Drossart}, {Eccleston},
  {Hartogh}, {Heske}, {Leconte}, {Micela}, {Ollivier}, {Pilbratt}, {Puig},
  {Turrini}, {Vandenbussche}, {Wolkenberg}, {Beaulieu}, {Buchave}, {Ferus},
  {Griffin}, {Guedel}, {Justtanont}, {Lagage}, {Machado}, {Malaguti}, {Min},
  {N{\o}rgaard-Nielsen}, {Rataj}, {Ray}, {Ribas}, {Swain}, {Szabo}, {Werner},
  {Barstow}, {Burleigh}, {Cho}, {du Foresto}, {Coustenis}, {Decin}, {Encrenaz},
  {Galand }, {Gillon}, {Helled}, {Morales}, {Mu{\~n}oz}, {Moneti}, {Pagano},
  {Pascale}, {Piccioni}, {Pinfield}, {Sarkar}, {Selsis}, {Tennyson}, {Triaud},
  {Venot}, {Waldmann}, {Waltham}, {Wright}, {Amiaux}, {Augu{\`e}res},
  {Berth{\'e}}, {Bezawada}, {Bishop}, {Bowles}, {Coffey}, {Colom{\'e}},
  {Crook}, {Crouzet}, {Da Peppo}, {Sanz}, {Focardi}, {Frericks}, {Hunt},
  {Kohley}, {Middleton}, {Morgante}, {Ottensamer}, {Pace}, {Pearson},
  {Stamper}, {Symonds}, {Rengel}, {Renotte}, {Ade}, {Affer}, {Alard}, {Allard},
  {Altieri}, {Andr{\'e}}, {Arena}, {Argyriou}, {Aylward}, {Baccani}, {Bakos},
  {Banaszkiewicz}, {Barlow}, {Batista}, {Bellucci}, {Benatti}, {Bernardi},
  {B{\'e}zard}, {Blecka}, {Bolmont}, {Bonfond}, {Bonito}, {Bonomo}, {Brucato},
  {Brun}, {Bryson}, {Bujwan}, {Casewell}, {Charnay}, {Pestellini}, {Chen},
  {Ciaravella}, {Claudi}, {Cl{\'e}dassou}, {Damasso}, {Damiano}, {Danielski},
  {Deroo}, {Di Giorgio}, {Dominik}, {Doublier}, {Doyle}, {Doyon}, {Drummond},
  {Duong}, {Eales}, {Edwards}, {Farina}, {Flaccomio}, {Fletcher}, {Forget},
  {Fossey}, {Fr{\"a}nz}, {Fujii}, {Garc{\'\i}a-Piquer}, {Gear}, {Geoffray},
  {G{\'e}rard}, {Gesa}, {Gomez}, {Graczyk}, {Griffith}, {Grodent}, {Guarcello},
  {Gustin}, {Hamano}, {Hargrave}, {Hello}, {Heng}, {Herrero}, {Hornstrup},
  {Hubert}, {Ida}, {Ikoma}, {Iro}, {Irwin}, {Jarchow}, {Jaubert}, {Jones},
  {Julien}, {Kameda}, {Kerschbaum}, {Kervella}, {Koskinen}, {Krijger}, {Krupp},
  {Lafarga}, {Landini}, {Lellouch}, {Leto}, {Luntzer}, {Rank-L{\"u}ftinger},
  {Maggio}, {Maldonado}, {Maillard}, {Mall}, {Marquette}, {Mathis}, {Maxted},
  {Matsuo}, {Medvedev}, {Miguel}, {Minier}, {Morello}, {Mura}, {Narita},
  {Nascimbeni}, {Nguyen Tong}, {Noce}, {Oliva}, {Palle}, {Palmer}, {Pancrazzi},
  {Papageorgiou}, {Parmentier}, {Perger}, {Petralia}, {Pezzuto},
  {Pierrehumbert}, {Pillitteri}, {Piotto}, {Pisano}, {Prisinzano}, {Radioti},
  {R{\'e}ess}, {Rezac}, {Rocchetto}, {Rosich}, {Sanna}, {Santerne}, {Savini},
  {Scandariato}, {Sicardy}, {Sierra}, {Sindoni}, {Skup}, {Snellen}, {Sobiecki},
  {Soret}, {Sozzetti}, {Stiepen}, {Strugarek}, {Taylor}, {Taylor}, {Terenzi},
  {Tessenyi}, {Tsiaras}, {Tucker}, {Valencia}, {Vasisht}, {Vazan}, {Vilardell},
  {Vinatier}, {Viti}, {Waters}, {Wawer}, {Wawrzaszek}, {Whitworth}, {Yung},
  {Yurchenko}, {Osorio}, {Zellem}, {Zingales}, \&
  {Zwart}}]{2018ExA....46..135T}
{Tinetti}, G., {Drossart}, P., {Eccleston}, P., {et~al.} 2018, Experimental
  Astronomy, 46, 135

\bibitem[{Verne(1865)}]{Verne}
Verne, J. 1865, De la terre {\`a} la lune (Pierre-Jules Hetzel)

\end{thebibliography}
\bibliographystyle{aa}

\end{document}